\documentclass{article}

     \PassOptionsToPackage{numbers, compress}{natbib}

     \usepackage[final]{neurips_2019}

\usepackage[utf8]{inputenc} 
\usepackage[T1]{fontenc}    
\usepackage{hyperref}       
\usepackage{url}            
\usepackage{booktabs}       
\usepackage{amsfonts}       
\usepackage{nicefrac}       
\usepackage{microtype}      
\usepackage{graphicx}
\title{Dual Neural Network Architecture for Determining Epistemic and Aleatoric Uncertainties}

\author{%
  Augustin ~Prado\thanks{work done during internship at Schlumberger-Doll Research}, \hspace{0.005cm} Ravinath Kausik\footnotemark[2], \hspace{0.005cm} Lalitha Venkataramanan\thanks{corresponding author}\\
    Schlumberger-Doll Research \\
  1 Hampshire Street, Cambridge, MA, 02139, USA\\
  \texttt{augustin.prado@epfl.ch, \{rviswanathan, lvenkataramanan\}}@slb.com \\
}

\begin{document}

\maketitle


\section{Introduction}

Deep learning techniques have been shown to be extremely effective for various classification and regression problems, but quantifying the uncertainty of their predictions and separating them into the epistemic and aleatoric fractions is still considered challenging. In oil and gas exploration projects, tools consisting of seismic, sonic, magnetic resonance, resistivity, dielectric and/or nuclear sensors are sent downhole through boreholes to probe the earth’s rock and fluid properties. The measurements from these tools are used to build reservoir models that are subsequently used for estimation and optimization of hydrocarbon production. Machine learning algorithms are often used to estimate the rock and fluid properties from the measured downhole data. Quantifying uncertainties of these properties is crucial for rock and fluid evaluation and subsequent reservoir optimization and production decisions.  These machine learning algorithms are often trained on a ‘ground-truth’ or core database. During the inference phase which involves application of these algorithms to field data, it is critical that the machine learning algorithm flag data as ‘out of distribution’ from new geologies that the model was not trained upon. It is also highly important to be sensitive to heteroscedastic aleatoric noise in the feature space arising from the combination of tool and geological conditions. Understanding the source of the uncertainty and reducing them is key to designing intelligent tools and applications such as automated log interpretation answer products for exploration and field development. In this paper we describe a methodology consisting of a system of dual networks comprising of the combination of a Bayesian Neural Network (BNN) and an Artificial Neural Network (ANN) addressing this challenge for geophysical applications. 

BNNs have been traditionally used for epistemic uncertainty estimation in supervised learning settings, for both regression and classification problems. This method works by determining the posterior weight distributions of the neural networks, providing us with both a mean and variance of the estimated outputs using variational inference. For these reasons, BNNs can be used to differentiate in and out-of-distribution predictions [1], thereby identifying the suitability of the trained models for application in new geological formations. On the other hand, aleatoric uncertainty is due to the noise inherent in the observations and can be heteroscedastic. The non-stationary nature of noise is due to properties of the geology and the response of different sensors to environmental factors. Recently a few different methods have been introduced to assign heteroscedastic data dependent standard deviations to outputs of neural networks. Gal et al [1], do this by minimizing a cost function with respect to the model parameters $\theta$ and variance $\sigma_i^2$ corresponding to output $y_i$,
\begin{center}
  $$
  L(x,y,\theta)  = \frac{1}{N} \sum_{i=1}^{n} \frac{1}{\sigma_i^{2}} \mid\mid f_\theta{(x_i)}-y_i \mid\mid^2 + \log[\sigma_i^{2}] \eqno{(1)}
  $$
\end{center}

The first term represents the accuracy of the model while the second term prevents the assignment of very high uncertainty values for all data points. Another method that has been recently introduced, is the \textit{deep direct estimation method} [2] which proposes the application of two neural networks. One of the ANNs is used approximate the conditional mean while the other ANN is used to estimate the conditional point wise variance respectively. However, a separation of the epistemic and aleatoric uncertainty cannot be made. 

\section{Dual Network approach}

In this paper we demonstrate a methodology consisting of a dual network system of a BNN and an ANN, for determining the total uncertainty and it’s separation into the epistemic and aleatoric fractions. The schematic representation of the dual network workflow is shown in figure 1. The details of the workflow are as follows. The probabilistic Bayesian Neural Network (BNN) is trained to estimate the mean and variance of the output using variational inference, with its weights being assumed to be Gaussian. The errors in the prediction of the BNN ($\hat{y}_i - y_{true})$) are fed into a second ANN. The goal of the second ANN network is to estimate the total variance using \textit{direct estimation method} [2], trained to correlate the features to the mean square error of the first BNN, as shown in equation 2. Both networks are trained simultaneously and therefore optimized together.   

\begin{figure}
	\centering
	\fbox{\rule[0cm]{0cm}{.1cm} \rule[0cm]{.1cm}{0cm}\includegraphics[width=0.5\linewidth]{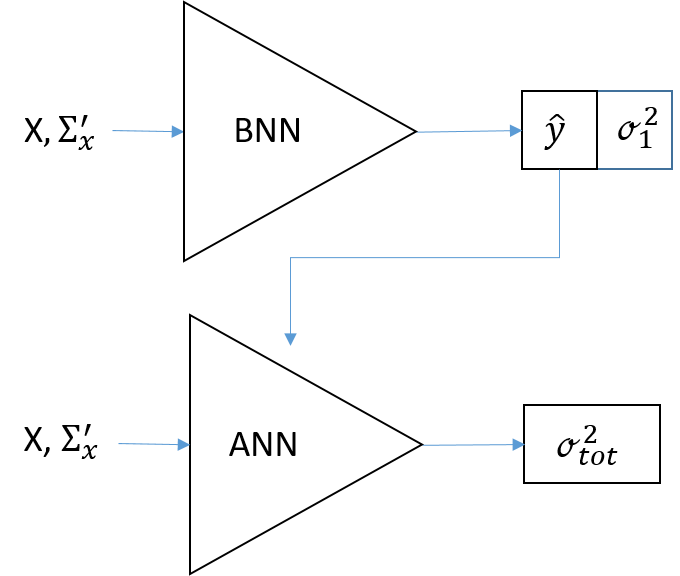}} 
	\caption{The dual network system consisting of a BNN and ANN is shown above. The BNN is used to estimate a mean and variance (epistemic uncertainty) of the output. The mean output of the BNN is input into a second network (ANN) for the determination of the total uncertainty ($\sigma_{\rm tot}$). The difference between this uncertainty and that of the BNN ($\sigma_1$) is a reflection of the aleatoric uncertainty of the data.}
\end{figure}

 \begin{center}
	
	$$
	\min_{\theta}\sum_{i=1}^{n}[h_\theta(x_i,y_{true})-(\hat{y}_i - y_{true})^2]^2 \eqno{(2)}
	$$
	
\end{center}

While the BNN estimates the mean value of the output ($\hat{y}$) and the epistemic uncertainty $\sigma_1$, the second network outputs the total uncertainty. The difference enables us to have an idea of the aleatoric uncertainty ($\sigma_2$) as shown below.

$$
 \sigma_{tot}^2 = \sigma_1^2 + \sigma_2^2\eqno{(3)}
$$

The total uncertainty can then be calibrated to get better results for diverse training datasets [3]. One of the key benefits of this approach is a better understanding of the components of the uncertainty, as the epistemic uncertainty is a good metric to differentiate in and out-of-distribution datasets (especially valuable for testing formation evaluation models in different geological formations), while the aleatoric uncertainty is sensitive to the heteroscedastic noise inherent to the data, within the training feature space. Understanding the aleatoric uncertainty originating from the heteroscedastic noise in the feature space can also help with experimental design and to optimize sampling schemes. This workflow therefore is a more complete method for deep learning approaches for geophysical applications. 

\begin{figure}
	\centering
	\fbox{\rule[0cm]{0cm}{.1cm} \rule[0cm]{.1cm}{0cm}\includegraphics[width=0.95\linewidth]{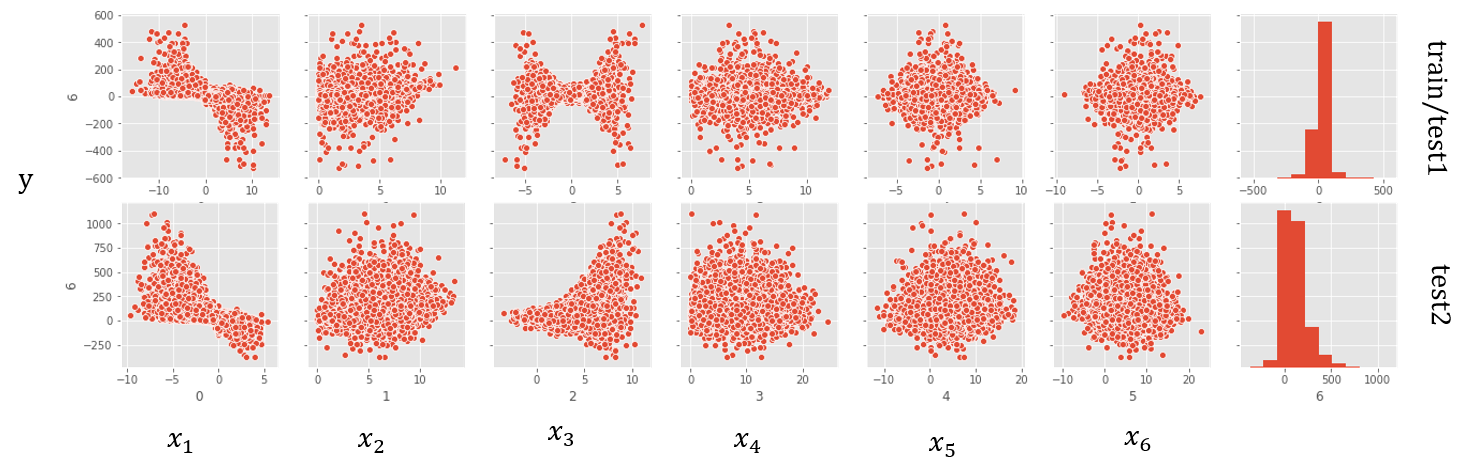}} 
	\caption{Pair plot of the input features of the test, and train 1 with in-distribution features and train 2 with out-of-distribution features is shown above.  }
\end{figure}

\subsection{Demonstration on synthetic data}

We demonstrate the application of this workflow on a synthetic dataset, below. A training dataset of 1000 points was created with input values $x_i$ (i = 1 to 6), sampled from Gaussian distributions with mean values of 0,3,0,5,0,0 and standard deviations of 4,2,2,2,2,2, respectively. Test datasets 1 and 2 were created with in-distribution data which has input features following the same distribution as the training data, and out-of-distribution data with features having only a mild overlap with the training set, respectively. The training dataset was made to satisfy the following relation,

$$
y_i = \sin(x_1) + x_2^2 -2x_1x_3^2 + \sqrt{x_4} + \exp(-x_2x_5^2) - \frac{3x_6}{0.2 + abs(x_1)} + \epsilon \eqno{(4)}
$$
with $\epsilon \sim N(0,\sigma_{noise})$ and $\sigma_{noise} \sim [0,2]$.
A dual network system was used to train on this dataset, and to test on in-distribution and out-of-distribution data. Both networks (BNN and ANN), had a similar architecture of 3 hidden layers with 20 neurons each. Tanh activation function was used for the BNN’s hidden layers and ReLU(x) = max(0,x) for ANN’s hidden layers. The results from the first part of the workflow, namely the Bayesian Neural Network is shown below in figure 3. The first plot is the learning curve showing the root-mean-square-error (RMSE) as a function of iterations, reflecting the accuracy of the network.  The second and third plots show the correlation of the epistemic (model) uncertainty with the absolute error for training, in-distribution and out-of-distribution datasets.

 \begin{figure}
 	\centering
 	\fbox{\rule[0cm]{0cm}{.1cm} \rule[0cm]{.1cm}{0cm}\includegraphics[width=0.95\linewidth]{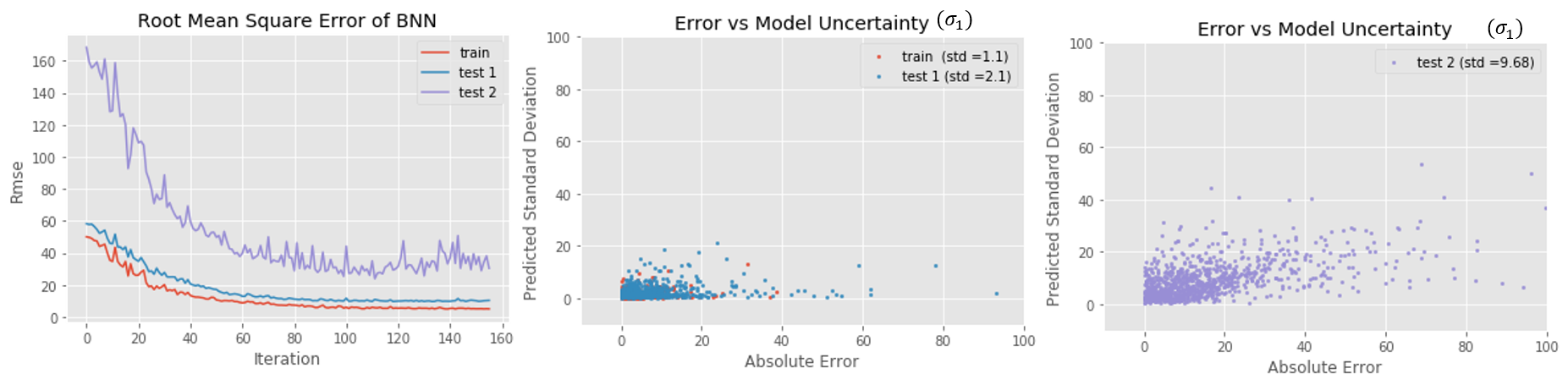}} 
 	\caption{The left plot shows the RMSE of the BNN as a function of iteration number, reflecting the higher accuracy of the training and in-distribution test data in comparison to the out-of-distribution test data. The correlation of the epistemic uncertainty with the absolute error are shown in the middle and right plots respectively.}
 \end{figure}

It can be inferred from the figure above that the uncertainties determined by the Bayesian Neural Network namely $\sigma_1$ (epistemic), are higher for the out-of-distribution data sets in comparison to the in-distribution datasets. Additionally the BNN uncertainty $\sigma_1$ is not well correlated with absolute error. The results from the second part of the workflow, the ANN, that estimates the total uncertainty is shown in figure 4. The first plot shows the root mean square error (RMSE) of the prediction of total variance for training, in-distribution and out-of-distribution data. The second and third plots compare the estimated aleatoric uncertainties (determined from the ANN output of total conditional variance $\sigma_{tot}$ using \textit{direct estimation} and the \textit{Law of total variance} to obtain aleatoric uncertainty $\sigma_2$ as shown in equation 3) versus the absolute error for the training, in-distribution and out-of-distribution datasets. It can be seen from figure 4 that the aleatoric uncertainty $\sigma_2$  is well correlated to the absolute error. The second network could also be converted into a BNN or a dropout-ANN network to obtain the variance of total uncertainty.
 
 \begin{figure}
	\centering
	\fbox{\rule[0cm]{0cm}{.1cm} \rule[0cm]{.1cm}{0cm}\includegraphics[width=0.95\linewidth]{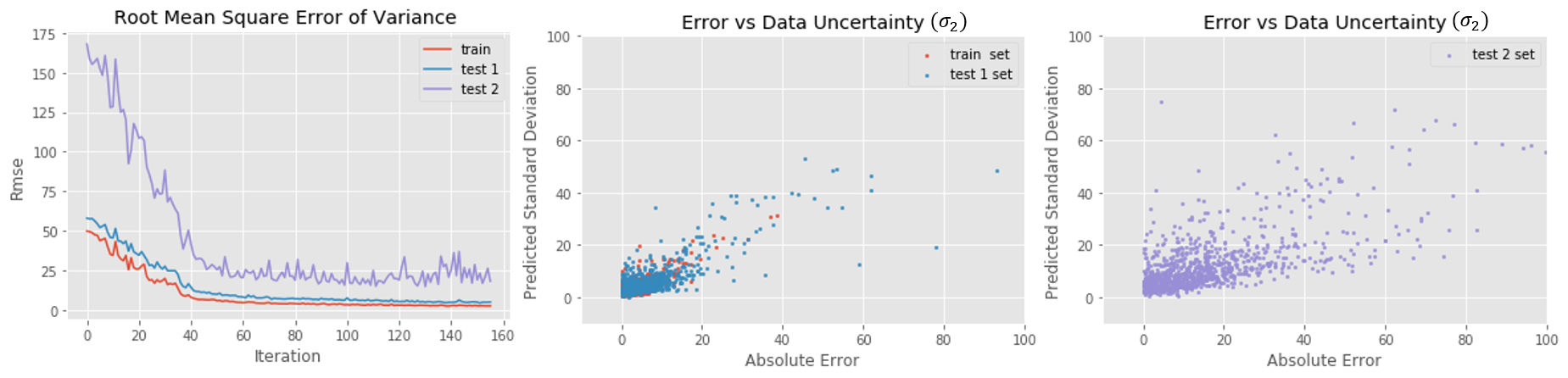}} 
	\caption{The left plot shows the RMSE for the total uncertainty predicted by the ANN as a function of iteration number, reflecting the higher accuracy of the training and ID test data in comparison to the out-of-distribution test data. The correlation of the aleatoric uncertainties with the absolute error for the training, in-distribution and out-of-distribution data are shown in the middle and right plots.}
\end{figure}

A demonstration of the impact of heteroscedastic noise on the datasets is shown below in figure 5. Heteroscedastic noise of $\sigma_i=1$ for $x_1 < 0$ else $\sigma_i=5$ (low and high noise conditions respectively) was added to the dataset. The epistemic and aleatoric uncertainties for the training datasets with low noise and high noise is displayed in figure 5. While the epistemic uncertainty does not separate the low and high noise datasets, the aleatoric uncertainty separates the two datasets, demonstrating the value of this workflow.   
 
 \begin{figure}
	\centering
	\fbox{\rule[0cm]{0cm}{.1cm} \rule[0cm]{.1cm}{0cm}\includegraphics[width=0.95\linewidth]{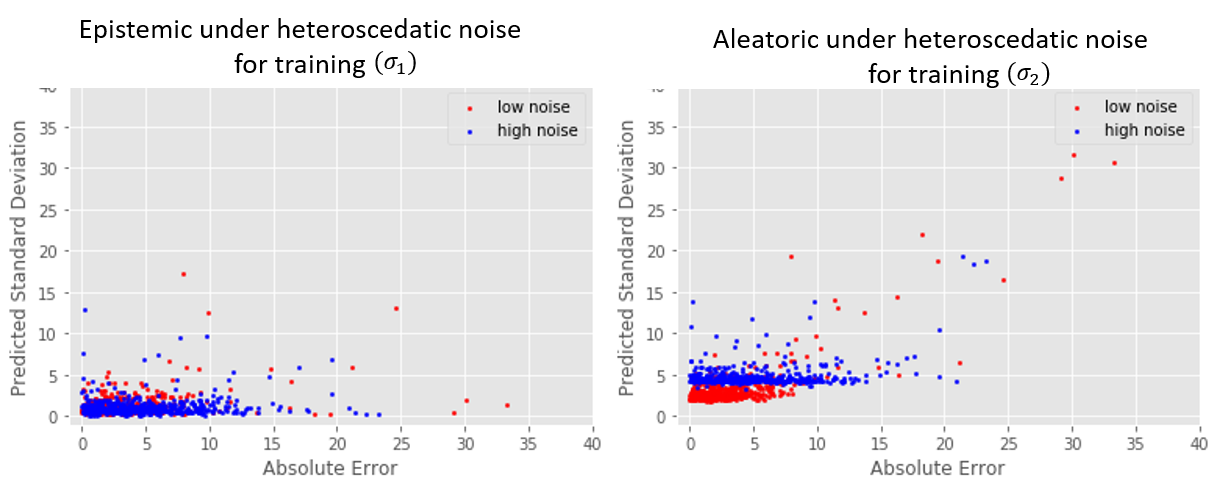}} 
	\caption{The ability of aleatoric uncertainty to separate the datasets based on the inherent noise is shown above.}
\end{figure}

\section{Summary}
In formation evaluation applications, it is important to understand and quantify the uncertainty in geophysical properties. The contribution of uncertainty has two factors: aleoteric and epistemic. Aleoteric uncertainty can be heteroschedastic in nature due to properties of the geological formation and response of measurements to different environmental factors. We propose a framework consisting of  a dual networks for inferring a geophysical property and its associated epistemic and aleoteric uncertainty. We have demonstrated that when input features are out-of-distribution, the output epistemic uncertainty is high. When the input features have high uncertainty, the aleoteric uncertainty is correspondingly higher. The workflow has been developed and demonstrated on simulated data sets with promising results. This workflow is currently being used for various geophysical applications.

\subsubsection*{Acknowledgments}

The authors are grateful to Schlumberger for supporting the project.

\section*{References}
\medskip

\small

[1] Kendall, A\ \& Gal, Y.\ (2017) What uncertainties do we need in Bayesian deep learning for computer vision? {\it Advances in neural information processing systems }

[2] Jonas, A\ \& Ktem, O.\ (2018) Deep bayesian inversion. {\it arXiv preprint arXiv:1811.05910 } 

[3] Volodymyr, K., Fenner, N.\ \& Ermon, S.\ (2018) Accurate uncertainties for deep learning using calibrated regression {\it arXiv preprint arXiv:1807.00263}

\end{document}